\documentclass[amsmath,amssymb,nofootinbib,notitlepage,superscriptaddress,twocolumn]{revtex4-1}
\pdfoutput=1

\usepackage{aas_macros,mathtools,esint,graphicx,overpic,xcolor,multirow,tabularx}
\usepackage[bottom]{footmisc}
\usepackage[colorlinks=true]{hyperref}
\definecolor{red}{RGB}{228,26,28}
\definecolor{green}{RGB}{77,175,74}
\definecolor{blue}{RGB}{55,126,184}
\definecolor{purple}{RGB}{152,78,163}

\let\originalleft\left
\let\originalright\right
\renewcommand{\left}{\mathopen{}\mathclose\bgroup\originalleft}
\renewcommand{\right}{\aftergroup\egroup\originalright}

\newcommand{\br}[1]{\left[#1\right]}
\newcommand{\cu}[1]{\left\{#1\right\}}
\newcommand{\pa}[1]{\left(#1\right)}

\newcommand{\pd}{\mathop{}\!\partial}

\newcommand{\ed}{\mathop{}\!\mathrm{d}}

\renewcommand{\O}[1]{\mathcal{O}\pa{#1}}

\begin{document}

\title{Signatures of particle collisions near extreme black holes}
\author{Delilah E.~A. Gates}
\email{deagates@princeton.edu}
\affiliation{Princeton Gravity Initiative, Princeton University, Princeton, New Jersey 08544, USA}

\author{Shahar~Hadar}
\email{shaharhadar@sci.haifa.ac.il}
\affiliation{
Department of Mathematics and Physics, University of Haifa at Oranim, Kiryat Tivon 3600600, Israel}
\affiliation{Haifa Research Center for Theoretical Physics and Astrophysics, University of Haifa, Haifa 3498838, Israel}

\begin{abstract}
Finite-energy particles in free fall can collide with diverging center-of-mass energy near rapidly rotating black holes. What are the most salient observational signatures of this remarkable geometric effect? Here we revisit the problem from the standpoint of the near-horizon extreme Kerr geometry, where these collisions naturally take place. It is shown that the ingoing particle kinematics admits a simple, universal form. Given a scattering cross section, determination of emission properties is reduced to evaluation of particular integrals on the sky of a near-horizon orbiting particle. We subsequently apply this scheme to the example of single-photon bremsstrahlung, substantiating past results which indicate that ejected particles are observable, but their energies are bounded by the rest masses of the colliding particles. Our framework is readily applicable for any scattering process.
\end{abstract}

\maketitle

\section{Introduction}
\label{sec:Introduction}

The rapidly-rotating, near-extreme Kerr geometry is a remarkable family of solutions of Einstein's equation describing a neutral black hole (BH) with nearly-maximal angular momentum. If such near-extreme BHs exist in nature, one of their notable properties, established in \cite{Piran1975,Banados2009,Harada2011}, is that free-falling particles can collide with parametrically large center-of-mass energy in their vicinity. From an asymptotic viewpoint, such collisions require a particular fine tuning of the particles' conserved energy and angular momenta, naturally realized when one particle is a near-horizon orbiter, while---importantly---these momenta are kept finite. 
Since a sizable portion of astrophysical BHs are believed to rotate quite rapidly \cite{Reynolds2019,Draghis2023}, this effect sparks the imagination as a possible avenue towards natural `celestial particle accelerators'. 

As the above-described collisions take place in the region just outside the BH event horizon, it is natural to question their detectability by an asymptotic observer; possible obstructions include infall of ejecta into the hole and parametrically large redshifts. Theoretical investigation of the collisions' observational signatures was pioneered in \cite{Piran1975a,Piran1977,Piran1977a}, where a particular focus was put on the process of inverse Compton scattering in a high-spin Kerr spacetime. 
The maximal energy of escaping photons was found to be of the order of the electron mass, i.e., in the gamma ray regime.
A related study \cite{Banados2011} considered the signatures of putative dark matter annihilation processes in the vicinity of extreme Kerr. There, the fraction of escaping particles from a single collision was analyzed for the particular case of escape in the equatorial plane.

Since \cite{Bardeen1999}, it was understood that the near-horizon, extremal Kerr (NHEK) limit is given by a simple, nondegenerate, symmetry-enhanced geometry. This insight inspired studies of possible implications for quantum BH physics \cite{Guica2009,Compere2012}, as well as potential observational signatures of high BH spin, e.g., \cite{Hadar2014,Gralla2017,Compere2017,Gates2018}.
It is no accident that the same NHEK geometry describes the arena in which such high-energy collisions take place.
In this paper, we revisit the problem of characterizing the collision signatures by employing a purely NHEK perspective. Our goal is to provide a simple, general framework to study the emission properties for any choice of scattering process.
Interestingly, taking the NHEK limit yields a unique, universal kinematic setup for ingoing particles which we derive in Sec.~\ref{sec:collisions}. We then develop the framework for computation of properties of (massless) particles emitted to asymptotic infinity in general scattering processes, working in the frame of a NHEK orbiter in which relativistic effects of escape/capture and red/blueshift are summarized by a simple, nontrivial geometric picture on the orbiter's sky \cite{Gates2021}, reviewed in Sec.~\ref{sec:universal kinematics}. We provide tools for efficient computation of the escape probabilities of ejecta and their red/blueshift, and of the expected energy; they are presented in Sec.~\ref{sec:escape general properties}. Subsequently, in Sec.~\ref{sec:brem}, we consider a particular example of a $2\to3$ process of single-photon bremsstrahlung. We first use momentum conservation to derive analytical bounds on the emitted and observed photon energies, and then use an approximate cross section for proton-electron bremsstrahlung to compute explicitly some characteristics of the emission in a particularly interesting range of collision energies.

Our results give a new perspective that corroborates the picture established in \cite{Piran1977a,Banados2011,Bejger2012,McWilliams2013,Harada2014,Leiderschneider2016}: signatures of NHEK high-energy collisions \emph{are} in principle observable; however, the energy of the escaping particles, 
in the processes that have been considered, are bounded by the rest mass of one of the colliding particles---namely, the NHEK orbiter. It would be interesting to consider more general scattering processes, especially with more particles involved, since 
this reduces the ratio of momentum constraints to degrees of freedom.
Our formalism
is suitable for the study of arbitrary NHEK processes, so it could be useful for contrasting different scattering cross sections and/or proving more general bounds. 
Herein we do not consider any possible astrophysical bounds on BH spin \cite{Thorne1974,Berti2009} as we are interested in understanding what is possible \emph{in principle}, all the way up to extremality.

\section{High-energy particle collisions near extremality}
\label{sec:collisions}

Neutral BHs of mass $M$ and angular momentum $J=aM$ are described by the Kerr metric, given in Boyer-Lindquist coordinates by\footnote{We use natural units $G_N=c=1$.}
\begin{align}\label{Kerr metric}
    ds^2=&-\frac{\Delta}{\Sigma}\left(dt-a\sin^2\theta  d\phi\right)^2+\frac{\Sigma}{\Delta}dr^2+\Sigma d\theta^2\nonumber\\
    &+\frac{\sin^2\theta}{\Sigma}\br{\left(r^2+a^2\right)d\phi-adt}^2 \,,
\end{align}
where $\Delta=r^2-2Mr+a^2$ and $\Sigma=r^2+a^2\cos^2\theta$.
Geodesic particle trajectories
in the Kerr geometry admit four independent integrals of motion which render geodesic motion in Kerr integrable: the particle mass $m$, its energy $E$, azimuthal angular momentum $L$, and Carter constant $Q=p^2_\theta-\cos^2\theta\left[ a^2(p^2_t-m^2)-p^2_\phi \csc^2\theta \right]$, where $p_\nu$ is the four-momentum. 
Conservation of $E$ and $L$ is guaranteed by stationarity and axisymmetry, respectively, and that of $Q$ derives from the existence of a rank-two Killing tensor.
In terms of these conserved quantities, the momentum of a geodesic particle in the Kerr geometry is given by
\begin{align}
	\label{eq:KerrMomentum}
	p_\nu\ed x^\nu=-E\ed t\pm_r\frac{\sqrt{\mathcal{R}(r)}}{\Delta(r)}\ed r\pm_\theta\sqrt{\Theta(\theta)}\ed\theta+L\ed\phi \,,
\end{align}
where 
\begin{align}
    \label{eq:RadialPotential}
	\mathcal{R}(r)=&\br{E\pa{r^2+a^2}-aL}^2\nonumber\\
	&-\Delta(r)\br{Q+\pa{L-aE}^2+m^2r^2},\\
	\label{eq:AngularPotential}
	\Theta(\theta)=&Q+a^2\pa{E^2-m^2}\cos^2{\theta}-L^2\cot^2{\theta},
\end{align}
play the role of effective potentials for radial and polar motion, respectively.

It is well-known \cite{Piran1975,Banados2009,Harada2011} that near-extremal, rapidly rotating Kerr BHs with $a=M\sqrt{1-\kappa^2}$ where $\kappa \ll 1$, act as `natural particle accelerators': finite-energy geodesic particles can collide with arbitrarily large center-of-mass energy near the BH horizon $r_+=M+\sqrt{M^2-a^2} = M(1+\kappa)$ when their angular momentum is properly tuned. 
The effect is naturally explained in terms of the near-horizon geometry \cite{Bardeen1999,Gralla2016}. In order to faithfully describe the latter, it is instrumental to introduce a 1-parameter family of coordinate transformations \cite{Bardeen1999,Hadar2014}
\begin{align}
    t=\frac{2MT}{\kappa^p}\,,~~r=r_+\left( 1+\kappa^p R \right)\,,~~\phi=\Phi+\frac{T}{\kappa^p}\,,
\end{align}
and take $\kappa\to0$ in \eqref{Kerr metric}. This procedure shows that at extremality the near-horizon region assumes a nondegenerate geometry with enhanced symmetry. For $0<p<1$, it yields the \emph{near horizon extreme Kerr (NHEK)} geometry
\begin{align}
\begin{gathered}
    \frac{ds^2}{2M^2 \Gamma }=-R^2dT^2+\frac{dR^2}{R^2}+d\theta^2+\Lambda^2\left(d\Phi+RdT\right)^2\,,\\
    \Gamma=\frac{1+\cos^2\theta}{2}, \quad \Lambda=\frac{2\sin\theta}{1+\cos^2\theta}, 
\end{gathered}
\end{align}
while for $p=1$, the so-called near-NHEK metric \cite{Bredberg2010} is obtained. 
$p$ determines the rate of scaling into the near-horizon region as the $\kappa \to 0$ limit is taken. For example, the innermost stable circular orbit (ISCO) radius scales with $p=2/3$ when $\kappa \to 0$, while the innermost (unstable) spherical photon orbit radius scales with $p=1$. 

Notably, timelike geodesics which are tuned to the superradiant bound $E=\Omega_H L$, where $\Omega_H=1/(2M)+\mathcal{O(\kappa)}$ is the horizon's angular velocity, can spend a long proper time in the NHEK, and are at parametrically large boost with respect to generic, non-tuned geodesics which promptly traverse the NHEK. This geometric feature is responsible for the high-energy collisions under discussion.
It seems plausible that such fine tuning could naturally arise in an accretion disk that includes quasi-circular orbiters and extends into the NHEK. 
Here we consider, for simplicity, only equatorial particles, thought of as part of an equatorial, geometrically-thin accretion disk.

Therefore, we consider two types of massive particles:
type I particles which 
circularly orbit deep in the NHEK with near-superradiant bound momentum \cite{Gates2018},
\begin{align}\label{eq:P1}
    p_{\mathrm{I}} \approx \frac{2M}{\sqrt{3}} m_{\mathrm{I}} d\Phi \, ,
\end{align}
where $m_{\mathrm{I}}$ is the mass of the particle, to leading order in $\kappa$, and type II `generic' particles which decouple from the average accretion flow at larger radii, and carry non-tuned angular momentum. As they plunge through the NHEK, to leading order in $\kappa$, their momentum becomes\footnote{Here the angular momentum is bounded by $2ME>L$.} 
\begin{align}\label{eq:P2}
    p_{\mathrm{II}} \approx -\frac{2ME-L}{\kappa^p} \, d\left(T-\frac{1}{R} \right) \, .
\end{align}
The momenta $p_{(\mathrm{I/II})}$ are `attractors' in the sense that, without artificial fine tuning and before taking collisions into account, particle momentum in the throat can naturally tend to one of the values \eqref{eq:P1},\eqref{eq:P2}. Furthermore, the invariant collision energy $p_{\mathrm{I}}\cdot p_{\mathrm{II}}
\sim\kappa^{-p}$ generically diverges as $\kappa \to 0$. Indeed, the rest frames of particles I and II are related by a large boost.
Therefore, to understand the observable effects of these high-energy collisions in the NHEK, we will consider the \emph{universal kinematics} defined by \eqref{eq:P1},\eqref{eq:P2} as ingoing momenta for the colliding particles. It will be convenient to analyze the collision in the frame of the circular orbiter \cite{Gates2021}, particle I, and use the differential cross section for particle production to answer various questions on the observability of the process. For concreteness, in Sec.~\ref{sec:brem} we focus on single-photon production in collisions of two such incoming particles.

\section{Universal kinematics of collision in the sky of a NHEK orbiter} \label{sec:universal kinematics}

\begin{figure}
    \centering
	\includegraphics[width=.9\columnwidth]{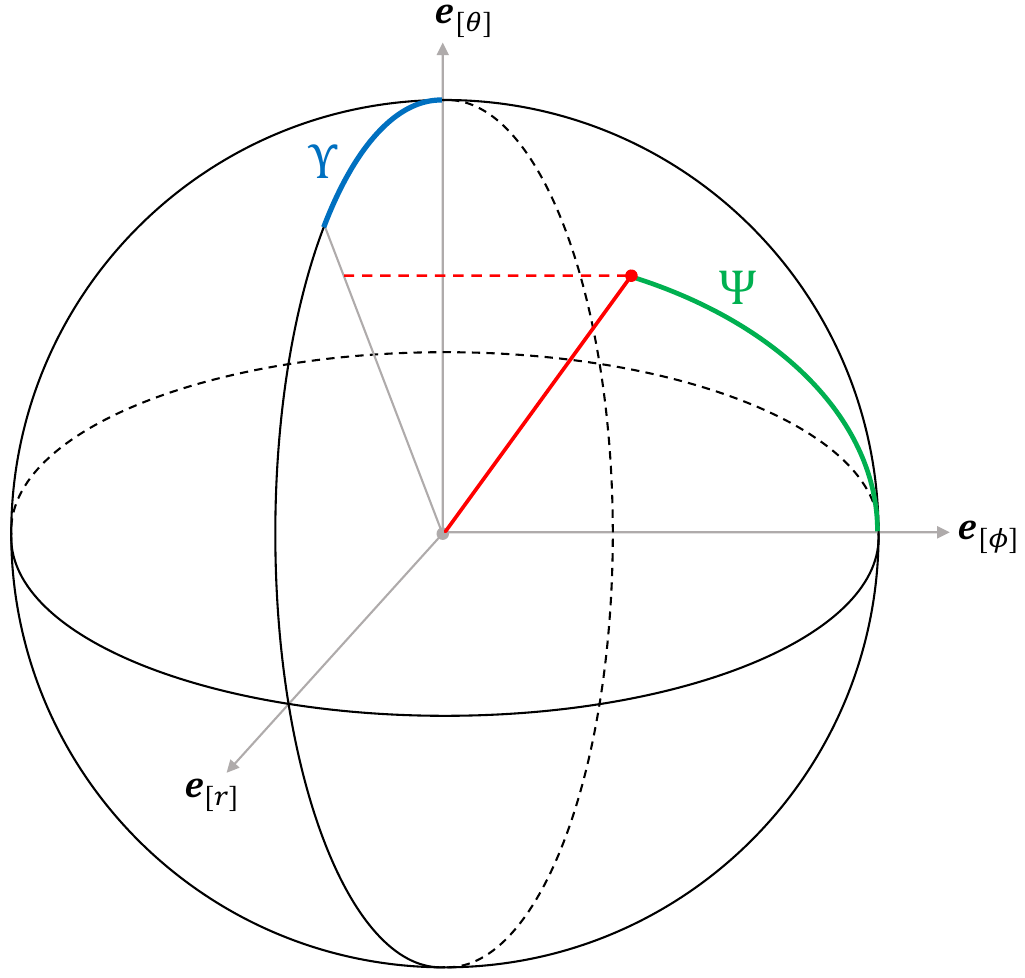}
	\caption{Angles parametrizing the orbiter sky. $\Psi$ is measured from the orbiter's direction of travel, i.e., the forward direction $\mathbf{e}_{[\phi]}$. $\Upsilon$ is measured from the direction perpendicular to the equatorial plane $\mathbf{e}_{[\theta]}$, in the plane perpendicular to the forward direction. $\mathbf{e}_{[r]}$ is the final (outwards) direction in the orbiter's orthnormal frame \eqref{eq:OrthornomalFrame}.}
	\label{fig:OrbiterSkys}
\end{figure}

We will analyze the collision in the frame of particle I, the circular orbiter.
An especially useful concept for the present analysis is the orbiter's \emph{sky}: a 2-sphere parameterizing spatial directions of emission/arrival of (null) geodesics at the orbiter's rest frame. The exact map between the sky of a general circular orbiter in Kerr and the integrals of motion $L/E$, $Q/E^2$ of the corresponding (null) geodesics is reviewed in App.~\ref{app:CircularOrbitersFrame}. 
Points on the sky are labeled by $\Psi \in \br{0,\pi}$, the polar angle measured from the orbiter's direction of motion,
and $\Upsilon \in \left(-\pi,\pi \right]$, the azimuthal angle measured from the frame axis parallel to the BH spin axis,
as illustrated in Fig.\ref{fig:OrbiterSkys}.

To deduce the properties of potentially observable photons produced by the collisions, we need to compute a few special objects and properties of the sky of particle I. First, using \eqref{eq:P2} and \eqref{eq:SkyAngles} in the near-horizon, near-extremal limit, the sky angle corresponding to the direction of motion of the incoming particle II is
\begin{align}    \label{eq:CollisionCenter}\pa{\Psi_{\mathrm{II}},\Upsilon_{\mathrm{II}}}=\pa{\frac{2\pi}{3},-\frac{\pi}{2}}.
\end{align}
The cross section is invariant under rotations which leave \eqref{eq:CollisionCenter} fixed.
The angle $\theta$ between \eqref{eq:CollisionCenter} and any direction to which an outgoing photon is emitted $(\Psi,\Upsilon)$ is given by
\begin{align}
    \label{eq:OutgoingEmissionAngle}
    \cos\theta=-\frac{\sqrt{3}}{2} \sin\Psi \sin\Upsilon - \frac{1}{2} \cos\Psi.
\end{align}

Second, we need the \emph{critical curve}, a closed curve on the sky which delineates the directions of capture inside the BH from those of escape to asymptotic infinity for emitted massless particles. 
Remarkably, in the near-horizon, near-extremal limit, the curve assumes a simple, universal fixed-point value, which is independent of the orbiter's NHEK radius and is given by $\left(\tilde{\Psi}(\Upsilon),\Upsilon\right)$, with
\begin{align}
    \label{eq:CriticalCurveNHEK}
	\cos{\tilde{\Psi}}(\Upsilon)=
	\begin{cases}
        \frac{\sqrt{\pa{3+\cos^2{\Upsilon}}\cos^2{\Upsilon}}-2}{4+\cos^2{\Upsilon}}
        &0\le \Upsilon \le\pi,\\
	    0&-\pi\le \Upsilon \le0.
	\end{cases}
\end{align}
Computing \eqref{eq:CriticalCurveNHEK} requires a novel type of near-horizon triple-scaling limit where one keeps track of the rates at which: a) the BH tends to extremality; and both b) the orbiter,  
and c) the photon shell radii \eqref{eq:PhotonShell} scale close to the horizon \cite{Gates2021}.  
The resulting critical curve is shown in Fig.~\ref{fig:CrossSectionRegions}. We will denote by $\mathcal{E}$ the region of the sky for which $\cos\Psi>\cos\tilde\Psi$, corresponding to photon escape. $\mathcal{E}$ covers $\approx54.64\%$ of the directions in the sky.

Third, we need the red/blueshift that escaping photons suffer/enjoy 
as they travel to the asymptotic region.\footnote{This quantity is well defined for all null geodesics, including those that fall into the horizon.} In terms of the emission angle, the red/blueshift factor is given by \cite{Gates2021}
\begin{align}
    \label{eq:Redshift}
    g(\Psi) = \frac{1}{\sqrt{3}} \, \left(1+2\cos\Psi\right).
\end{align}
For example, maximal blueshift occurs for forward emission, $g(\Psi=0)=\sqrt{3}$; critical emission, with $\cos\Psi\in\br{-1/2,0}$, has $g\in\br{0,1/
\sqrt{3}}$. We denote by $\mathcal{B}$ the sky region with $\cos\Psi>(\sqrt{3}-1)/2$, illustrated in Fig.~\ref{fig:CrossSectionRegions}. $\mathcal{B}$ corresponds to the blueshifted emission directions, all of which escape the BH; it covers $\approx31.70\%$ of the sky.

\section{Escape emission properties from a general scattering process} 
\label{sec:escape general properties}

With the universal kinematics and shape of the critical curve at hand, we can analyze properties of collision processes with photon emission as a function of their differential cross sections. In this paper we consider doubly differential cross sections $d\sigma/(dk d\Omega)$ only for single-photon emission, and integrate over all final states of the other outgoing particles. By symmetry, this cross section will be a function only of $k$, the photon energy/momentum in the rest frame of particle I, and of $\cos\theta$ \eqref{eq:OutgoingEmissionAngle}, the photon emission angle with respect to the direction of travel of particle II.
The probability of a photon to reach asymptotic null infinity given that it has energy $k$, the \emph{specific escape probability}, is given by
\begin{align}
    \mathcal{P}_\mathcal{E}(k) = \left(\frac{d\sigma}{dk}\right)^{-1} \int_{\mathcal{E}} d\Omega \, \frac{d\sigma}{dk d\Omega} \pa{k,\cos\theta}\,  ,
\end{align}
where $\mathcal{E}$ is the region of escape, and 
\begin{align}
    \frac{d\sigma}{dk}=\int_{S^2}d\Omega \, \frac{d\sigma}{dk d\Omega}\,,
\end{align}
is the total cross section at energy $k$. Similarly, the probability of a photon with energy $k$ to escape and reach infinity with $g>1$, the \emph{blueshifted specific escape probability}, is given by 
\begin{align}
    \mathcal{P}_\mathcal{B}(k) = \pa{\frac{d\sigma}{dk}}^{-1} \int_{\mathcal{B}} d\Omega \,\frac{d\sigma}{dk d\Omega} \,,
\end{align}
where $\mathcal{B}$ is the region of blueshifted emission.
The expectation value of the emitted energy per collision to infinity by a photon with energy $k$, the \emph{specific expected escape energy}, is given by
\begin{align}
    \mathcal{F}_\mathcal{E}(k) = \left(\frac{d\sigma}{dk}\right)^{-1}  \int_{\mathcal{E}} d\Omega\,\frac{d\sigma}{dk d\Omega}\,g(\Psi) \, k \, .
\end{align} 

\begin{figure}
    \centering \includegraphics[width=\columnwidth]{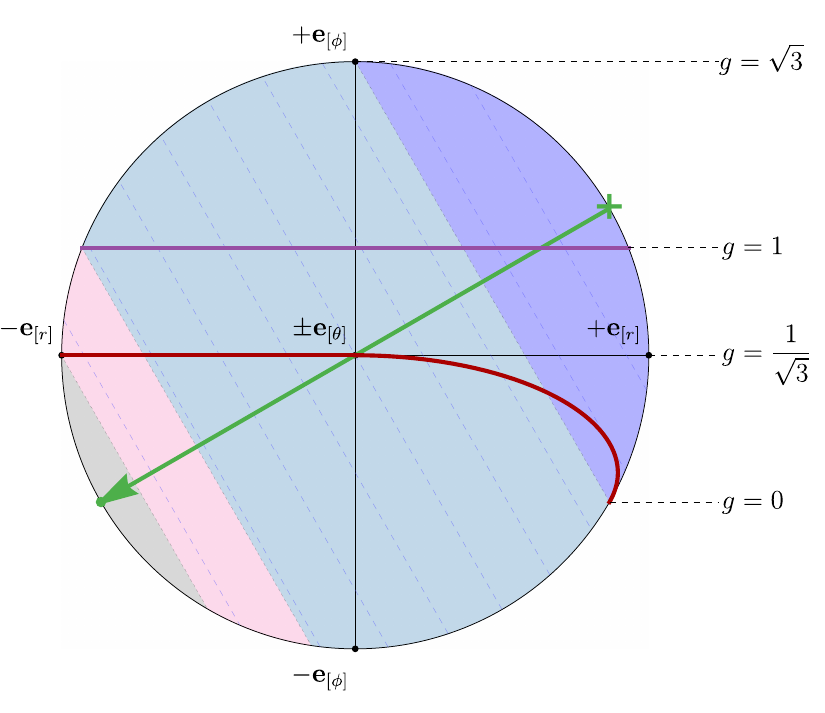}
	\caption{Sky of a NHEK orbiter, particle I, shown from above the $\mathbf{e}_{[\theta]}$ axis and marked by: the direction of motion of particle II \eqref{eq:CollisionCenter} (green arrow), lines of constant photon emission angle $\theta$ \eqref{eq:OutgoingEmissionAngle} (dashed blue lines),  
    the critical curve (red line) delineating directions of photon escape $\mathcal{E}$ (above) and capture $S^2\setminus\mathcal{E}$ (below), and directions of redshift factor unity (purple line)--above which lie directions of blueshifted escape $\mathcal{B}$. 
    The sky is partitioned into regions according to the value of $\theta$. $\theta_\mathrm{max}$ in the grey region implies the emitted photon cannot escape the BH; for $\theta_\mathrm{max}$ in the pink region, emitted photons may escape but only with $g<1$; when $\theta_\mathrm{max}$ is in the light blue or violet regions, photons may escape with $g>1$. When $\theta_\mathrm{max}$ lies in the violet region, the photon may escape with maximal blueshift $g=\sqrt{3}$.}
	\label{fig:CrossSectionRegions}
\end{figure}

In general, momentum conservation does not allow for photon emission in all directions $\theta\in\br{0,\pi}$ at any photon energy $k$. Instead, for certain values of $k$, the differential cross section is nonzero only for $\theta <\theta_\mathrm{max}(k)$, where $0\leq\theta_\mathrm{max}(k)\leq\pi$ depends on the process being considered. 
The consequences may be simply understood geometrically, as described below and illustrated in Fig.~\ref{fig:CrossSectionRegions}.

We partition the sky of particle I into regions corresponding to different ranges of values taken by $\theta(\Psi,\Upsilon)$. $\theta_\mathrm{max}(k)$ then determines which of these regions are accessible for photons produced with energy $k$, and therefore has implications for the escape probabilities, as follows:
\begin{itemize}
    \item $\cos\theta_\mathrm{max}\geq\frac{\sqrt{3}}{2} \Rightarrow \mathcal{P}_\mathcal{E}=0$, 
    \item $\frac{\sqrt{3}}{2} > \cos\theta_\mathrm{max}\geq \frac{1-\sqrt{3}+\sqrt{2} \ 3^{3/4}}{4} \Rightarrow \mathcal{P}_\mathcal{E}>\mathcal{P}_\mathcal{B}=0$,
    \item $\frac{1-\sqrt{3}+\sqrt{2} \ 3^{3/4}}{4} > \cos\theta_\mathrm{max} \, \Rightarrow \mathcal{P}_\mathcal{E}>\mathcal{P}_\mathcal{B}>0$.
\end{itemize}
Note that whenever $\cos\theta_\mathrm{max}\leq -1/2$, emission with maximal blueshift $g=\sqrt{3}$ can reach asymptotic null infinity.

We also define the \emph{total escape probability} and the \emph{total expected escape energy} at infinity from a single collision,
\begin{align} 
\label{eq:TotalProb}
    \Bar{\mathcal{P}}_\mathcal{E}&=\frac{1}{\sigma}\int_0^{k_*}dk\, \frac{d\sigma}{dk}\mathcal{P}_\mathcal{E}(k) =\frac{1}{\sigma}\int_0^{k_*} dk \,\int_{\mathcal{E}} d\Omega\,\frac{d\sigma}{dk d\Omega} \,,\\
\label{eq:TotalFlux}
    \Bar{\mathcal{F}}_\mathcal{E} &=\frac{1}{\sigma}\int_0^{k_*}dk\, \frac{d\sigma}{dk}\mathcal{F}_\mathcal{E}(k) =\frac{1}{\sigma}\int_0^{k_*} dk \,\int_{\mathcal{E}} d\Omega\,\frac{d\sigma}{dk d\Omega} \,g\,k\, ,
\end{align}
respectively, wherein
\begin{align}
    \sigma=\int_0^\infty dk\ \frac{d\sigma}{dk}  =\int_0^\infty dk \int_{S^2}  d\Omega \ \frac{d\sigma}{dk d\Omega} \ ,
\end{align}
is the total cross section area, and $k_*$ is the maximal photon escape energy 
defined via $\cos\theta_\mathrm{max}\pa{k_*}={\sqrt{3}}/{2}$.

Another interesting quantity we define is the \emph{specific expected observed energy} per collision, integrated over all observation angles, $f(k_\mathrm{obs})$. Exchanging the order of integration in \eqref{eq:TotalFlux}, and using \eqref{eq:Redshift} and $g=k_\mathrm{obs}/k$ where $k_\mathrm{obs}$ is the observed energy at infinity, we can write
\begin{align} \label{eq:FluxDef2}
    \sigma \bar{\mathcal{F}}_\mathcal{E}& = \int_0^{\sqrt{3}k_*} d k_\mathrm{obs}\int_{k_\mathrm{obs}/\sqrt{3}}^{k_*}  d k \ F\pa{k,k_\mathrm{obs}}\, \nonumber\\
    &=\int d k_\mathrm{obs} f(k_\mathrm{obs}) \, ,
\end{align}
where
\begin{align}
    F\pa{k,k_\mathrm{obs}}=\frac{\sqrt{3}k_\mathrm{obs}}{2k}\int_{\Upsilon_g^-( k_\mathrm{obs}/k)}^{\Upsilon_g^+( k_\mathrm{obs}/k)} d\Upsilon \frac{d\sigma}{dk d\Omega} \, ,
\end{align}
and
\begin{align}
    \Upsilon_g^\pm(g)=
\begin{cases}
    \arccos\pa{\pm\frac{2 g}{\sqrt{1+{2g}/{\sqrt{3}}-g^2}} }, \quad & 0<g<\frac{1}{\sqrt{3}} \,, \\
    \pm \pi, \quad & \frac{1}{\sqrt{3}}<g<\sqrt{3} \,.
\end{cases}    
\end{align}

\section{Example: bremsstrahlung} \label{sec:brem}

As a concrete demonstration, we will apply the general prescription outlined above to single-photon emission from 
bremsstrahlung. First we will use momentum conservation to derive general constraints on the observed emission, and then specialize to proton-electron bremsstrahlung (PEB)\footnote{Some sources in the literature distinguish PEB from electron-proton bremsstrahlung based on the kinematics---the second of the two listed particles is taken to be initially at rest. Here we refer to these processes, which are related by a boost, by the same term.} in the no-recoil approximation and explicitly compute properties of the emitted radiation.
Recall that we idealize to a geometrically thin, equatorial accretion disk of orbiting plasma composed of free electrons and nuclei, the lightest of which originate from ionized hydrogen and constitute of a single proton.

\begin{figure}
    \centering    \includegraphics[width=\columnwidth]{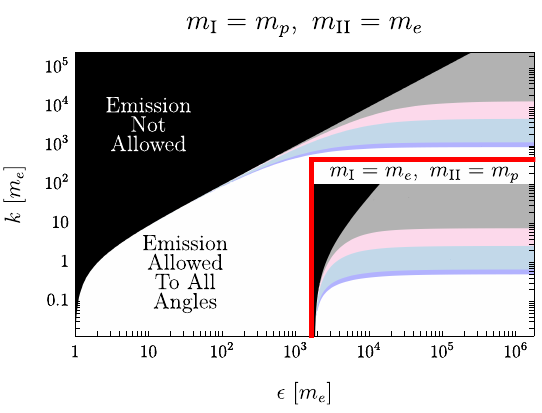}
	\caption{Momentum constraints on single-photon PEB between a NHEK orbiter, particle I, and an ingoing particle with `generic', untuned momentum, particle II. We partition the plot into regions where emission is disallowed (black), allowed to all directions in the orbiter sky (white), and allowed for $\theta<\theta_{\mathrm{max}}$ (grey, pink, light blue, violet, in accordance with the color scheme described in Fig.~\ref{fig:CrossSectionRegions}). The bulk of the plot shows the constrains when the orbiter is the proton; the inset in red shows the constraints when the orbiter is the electron.}
	\label{fig:BremPhaseSpace}
\end{figure}

For single-photon bremsstrahlung, momentum conservation implies
\begin{align} \label{eq:momentum conservation}
    k &= \frac{m_{\mathrm{I}} \epsilon + p'_{\mathrm{I}} \cdot p'_{\mathrm{II}}}{\epsilon+m_{\mathrm{I}}-\cos\theta\sqrt{\epsilon^2-m_{\mathrm{II}}^2}} \,,
\end{align}
where $m_{\mathrm{I/II}}$ are the masses of particles I/II, $\epsilon=p_{\mathrm{II}}^{[t]}$ is the energy of particle II in the rest frame of particle I, and $p'_{\mathrm{I/II}}$ are the momenta of the outgoing particles I/II. Since $p'_{\mathrm{I}} \cdot p'_{\mathrm{II}} \leq -m_{\mathrm{I}} m_{\mathrm{II}}$, \eqref{eq:momentum conservation} yields a maximal scattering angle for fixed $k$
\begin{align}
    \label{eq:CosThetaMaxPEB}
    \cos\theta_\mathrm{max}\pa{k}=\frac{1}{\sqrt{\epsilon^2-m_{\mathrm{II}}^2}}\pa{\epsilon+m_{\mathrm{I}}-\frac{m_{\mathrm{I}}(\epsilon-m_{\mathrm{II}})}{k}} \, .
\end{align}
When $\cos\theta_{\mathrm{max}}<-1$ emission is allowed to all sky directions.
In the ultrarelativistic limit $\epsilon \gg m_{\mathrm{II}}$,
\begin{align}
    \cos \theta_\mathrm{max} \approx 1-\frac{m_{\mathrm{I}}}{k} \,.
\end{align}
Thus, in high-energy single-photon bremsstrahlung, high-energy photons $k \gg m_{\mathrm{I}}$ are always beamed forward with $\theta_\mathrm{max}\approx0$, while photons of $k \sim m_{\mathrm{I}}$ may be emitted with larger deflection angles. Fig.~\ref{fig:BremPhaseSpace} graphically summarizes the momentum conservation constraints in the case of single-photon PEB.

For fixed $\theta$, \eqref{eq:momentum conservation} yields a maximal photon energy $k_\mathrm{max}$,
\begin{align}
    \label{eq:kmax}
    k_\mathrm{max} (\theta)=\frac{m_{\mathrm{I}}\pa{\epsilon-m_{\mathrm{II}}}}{\epsilon+m_{\mathrm{I}}-\cos\theta\sqrt{\epsilon^2-m_{\mathrm{II}}^2}} \,,
\end{align}
which is bounded above by the high-energy limit,
 \begin{align}
    \label{eq:kmax ultrarelativistic}
    \bar{k}(\theta)=
     \lim_{\epsilon\to\infty} k_{\mathrm{max}}(\theta)=\frac{m_{\mathrm{I}}}{1-\cos\theta}.
 \end{align}
Thus, the mass of particle II sets the maximal photon energy, while the particles' mass ratio controls the rate of approach of $k_\mathrm{max}$ to $\bar k$.
The maximal energy with which an escaping photon can be \emph{emitted} in the process under consideration is $k_* =\bar{k}\pa{\pi/6}\approx 7.46 m_{\mathrm{I}}$. 
A disparate attribute is the maximal \emph{observed} energy of an escaping photon $k_\mathrm{obs}=gk$, which is given by
\begin{align}
    \label{eq:MaxKobs}
    \max\limits_{(\Psi,\Upsilon)\in \mathcal{E}}{\left\{ g\pa{\Psi,\Upsilon}\bar{k}\left[\theta(\Psi,\Upsilon)\right]\right\} } \,.
\end{align}
It turns out that the maximum in \eqref{eq:MaxKobs} is obtained precisely on the critical curve at $(\Psi,\Upsilon)=(\pi/2,-\pi/2)$, and thus the observed photon energy is bounded by 
\begin{align}
    k^\mathrm{max}_\mathrm{obs}=\frac{k_*}{\sqrt{3}}=2\pa{1+\frac{2}{\sqrt{3}}}m_{\mathrm{I}}.
\end{align}

Keeping these general momentum constraints in mind, we now focus on the particular setup of single-photon PEB in the limit of negligible recoil. The no-recoil regime we focus on here exists thanks to the large mass ratio between electron and nucleus, which for ionized hydrogen is $m_p/m_e \approx 2\cdot 10^3$. In this range an especially simple analytic expression for the cross section is available, through use of the leading Born approximation, which was computed by Bethe and Heitler, and Sauter in 1934 \cite{Bethe1934,Sauter1934}. The explicit formula in the proton frame is given in App.~\ref{app: cross section formula}; cf. \cite{Koch1959,Haug2003} for more details. 
Note that while we have focused on PEB in the present paper, proton-proton and electron-electron scattering are perfectly viable processes to consider as well. When considering emission of energetic photons produced by such processes, recoil must be taken into account already for mildly relativistic collisions.

\begin{figure*}
    \centering
    \resizebox{\linewidth}{!}{
    \begin{tabular}{c c}
    \begin{tabular}{c c}
    \hspace{1cm}{\large\underline{${m_{\mathrm{I}}=m_p}, m_{\mathrm{II}}=m_e$}} & {\large\underline{$m_{\mathrm{I}}=m_e, \ m_{\mathrm{II}}=m_p$}}\hspace{1cm} \\[2.5ex]
    \includegraphics{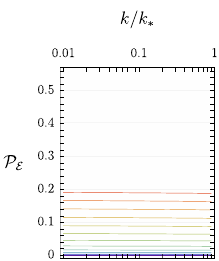}\quad & \includegraphics{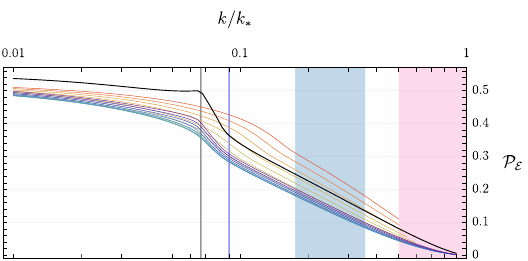}\\
    \includegraphics{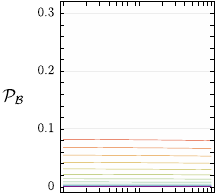}\ & \includegraphics{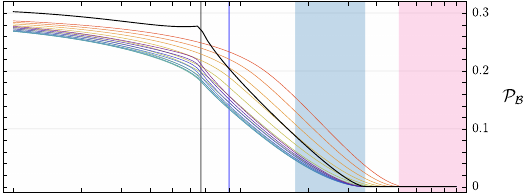}\\
    \includegraphics{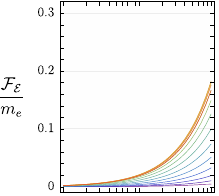}\ &  \includegraphics{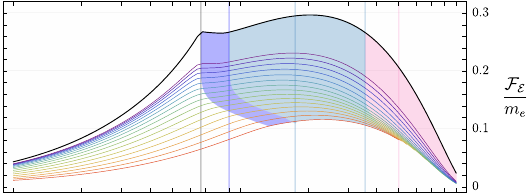}\\
    \includegraphics{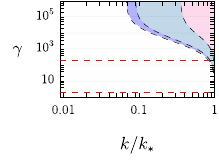}\ &  \includegraphics{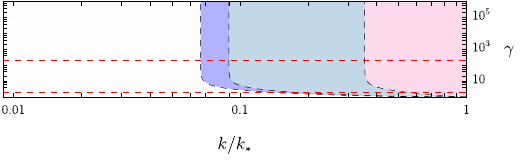}
    \end{tabular}&\quad
    \raisebox{-.47\height}{\includegraphics{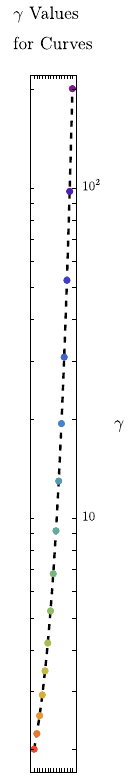}}
    \end{tabular}}
    \caption{Escaped emission properties of single-photon PEB between a NHEK orbiter, particle I, and a `generic' plunger, particle II, in the no-recoil approximation. We compare the proton-as-orbiter (left column) and the electron-as-orbiter (middle column) cases. The specific escape probability (top row), blueshifted specific escape probability (second row), and specific expected escape energy (third row) are presented as a function of $k/k_*$ at fixed $\gamma\in\br{2,200}$ (colored curves with explicit values shown on the right column). In the middle column the large $\gamma$ limit is indicated by the black curves. The fourth row is a recreation of Fig.~\ref{fig:BremPhaseSpace} in the region where emission may escape, upon which the region of interest $\gamma\in\br{2,200}$ is marked by red dashed lines. In this region, emission is kinematically allowed to (almost) all sky directions in the proton-as-orbiter case; nevertheless the radiation is significantly beamed towards directions of capture in the large $\gamma$ limit. In the electron-as-orbiter case, a significant range of photon emission energies $k$ have a kinematic cutoff on the sky emission angle $\theta_\mathrm{max}<\pi$.
    Notable regions where $\theta<\theta_\mathrm{max}$ are shaded using the color scheme established in Fig.~\ref{fig:CrossSectionRegions}. 
    For all $\gamma$, emission is allowed to the whole sky in the region left of the black line, and can reach infinity with maximal redshift factor $g=\sqrt{3}$ to the left of the violet line. Despite having a lower maximal escape photon energy $k_*$ and often less orbiter sky directions available, the electron-as-orbiter case generally has higher specific escape probabilities and expected escape energy, including nonzero values in the large $\gamma$ limit.}
    \label{fig:PEBStats}
\end{figure*}

Working in the rest frame of the orbiter, particle I, we explicitly compute the emission properties discussed in Sec.~\ref{sec:escape general properties} for the process $e^- p^+ \to e^- p^+ \gamma$ using the simple analytical expressions of \cite{Bethe1934,Sauter1934}; see Fig.~\ref{fig:PEBStats} for results.
The energy range we focus on in the numerical evaluation of the emission characteristics is $1\ll \gamma \ll m_p/m_e$, which is ultrarelativistic on one hand, but allows us to neglect the nucleus' recoil, or momentum transfer, on the other hand. 
We therefore consider $\gamma\in\br{2,200}$ in our computations. 
At even higher energies, the no-recoil approximation we use here breaks down.
We expect radiation in that case to be significantly more beamed towards directions of capture because of recoil, suppressing emission probabilities. Still, we calculate also the high-energy ($\gamma \to \infty$) limiting result (black solid curves in Fig.~\ref{fig:PEBStats}); while outside the no-recoil approximation for PEB, it provides a formal bound which heavier nuclei substituted for the proton can come closer to saturating. 
Another noteworthy point is that for $\cos\theta_\mathrm{max}>-1$, the no-recoil approximation breaks down for $\theta \gtrapprox \theta_\mathrm{max}(k)$ even in the energy range under consideration, since precise equality occurs when $p'_{p} \cdot p'_{e}=-m_{\mathrm{I}} m_{\mathrm{II}}$, implying significant momentum transfer.\footnote{Note that the PEB cross section in the approximation employed here is inapplicable, giving ill defined results, for $0.5\lesssim k/k_*$ when $\gamma \lesssim 7$ in the electron-as-orbiter case; we exclude this range in Fig.~\ref{fig:PEBStats}.}
Nevertheless, for simplicity, we use the no-recoil approximation all the way up to $\theta_\mathrm{max}(k)$, and for larger angles we directly enforce the vanishing of the cross section by multiplying it with a Heaviside theta function, $d\sigma/(dk d\Omega) \propto \Theta\br{\cos\theta-\cos\theta_\mathrm{max}(k)} $.

Concisely, we summarize our results for single-photon PEB in the no-recoil regime as follows. We find significant specific escape probability for photon emission at all energies up to a sizable fraction of $k_*$, corresponding to energies $\sim m_e$, deep in the gamma ray regime. Typically, the electron-as-orbiter case is more observable than the proton-as-orbiter case; often exhibiting comparable or larger specific escape probability and specific expected energy. When the electron is the orbiter, we observe a kink-like transition in the specific escape probability and specific expected energy that occurs when $\cos \theta_\mathrm{max}(k)=-1$; the kink becomes increasingly sharper in the $\epsilon \gg m_p$ regime. We now provide more details on the behavior of particular observables. \linebreak
\textbf{Specific escape probability and blueshifted specific escape probability.} When the NHEK orbiter (particle I) is a proton hit by an incoming electron (particle II), the escape probabilities for $\gamma=2$ are $\mathcal{P}_\mathcal{E}\lesssim0.2$ and $\mathcal{P}_\mathcal{B}\lesssim0.1$, respectively. As dictated by the momentum constraints, the maximal energy of the emitted photon is $\sim m_e$. The probabilities are nearly $k$-independent, but as may be expected, they rapidly approach zero as $\gamma$ is increased due to relativistic beaming. On the other hand, when the NHEK orbiter is an electron hit by an incoming proton, we find a significant probability of escape (sizable portion of unity) for photon emission at all energies up to a large fraction of $k_*$ even in the $1\ll\gamma$ limit. In fact, $0.3\lesssim\mathcal{P}_\mathcal{E}\lesssim 0.5$ and $0.15\lesssim\mathcal{P}_\mathcal{B}\lesssim 0.3$ for $k/k_*\lesssim0.1$. As mentioned above, we observe an interesting feature, a kink-like transition, in the escape probability that occurs when $\cos \theta_\mathrm{max}(k)=-1$; this kink becomes increasingly sharp in the $1 \ll \gamma$ regime. For photon energies larger than the kink energy, escape probabilities decrease rapidly with increasing $k$, as the escape region $\pa{\Psi,\Upsilon}\in\mathcal{E}$ for which $\cos\theta>\cos\theta_{\mathrm{max}}$ shrinks. (See first and second rows of Fig.~\ref{fig:PEBStats}.) \linebreak
\textbf{Specific expected escape energy.} 
When the NHEK orbiter is a proton, the expected energy grows as $\sim k^2$. The maximal expected escape energy is $\mathcal{F}_\mathcal{E}\approx0.2m_e$, achieved when $\gamma\sim\O{1}$ and $k\sim k_*$, and the expected escape energy decreases with increasing $\gamma$. When the NHEK orbiter is an electron, we find $\O{m_e}$ expected escape energy for a significant portion of photon energies $k$. The kink transition is also visible in expected escape energy, with $\mathcal{F_E}$ growing as $\sim k^2$ prior to the kink energy, peaking in the region where $-1/2<\cos\theta_\mathrm{max}<\pa{1-\sqrt{3}+\sqrt{2}\,3^{3/4}}/4$, and finally going to zero as $k$ approaches $k_*$. The peak in $\mathcal{F_E}$ appears after the kink, as the shrinkage of allowed escape directions competes with the photon energy increase. (See third row of Fig.~\ref{fig:PEBStats}.)

We conclude that, as in \cite{Piran1977a, Banados2011,Bejger2012,McWilliams2013,Harada2014,Leiderschneider2016}, single-photon PEB signatures of high-energy collisions can make it out to asymptotic infinity, but only with bounded-energy photons of $\sim m_e$.

It is instructive to note that the center-of-mass collision energy between an ISCO orbiter and an equatorial plunging particle is $\sim\kappa^{-1/3}$ \cite{Harada2011}.
Therefore, for ionized hydrogen the no-recoil approximation should hold up to $\kappa\sim10^{-9}$. 
Note also that we have not discussed neither the angular distribution of the radiation nor its optical appearance. The former would require a considerably more complicated analysis since the map between emission angle in the rest frame of particle I and angle of arrival at the celestial sphere is highly oscillatory for a NHEK emitter \cite{Gates2021}. Regarding the latter, we do know that a sufficiently inclined asymptotic observer will see most of the emitted photons showing up parametrically close to the NHEKline \cite{Gralla2017}. An exception is the neighborhood of the emission direction 
$\pa{{2\pi}/{3},{\pi}/{2}}$
which can show up far from the NHEKline, see App.~\ref{app:SuperradiantEmission}.

Finally we note that it would be interesting to investigate more general processes using our formalism. In particular, it seems conceivable that processes involving more emitted particles may sustain more freedom under the momentum constraints and be able to emit more energetic particles to infinity. The question of whether this is the case, or there exists a general bound on the energy of emitted particles, is left for future investigation.

\acknowledgements
We are grateful to Alejandro Cardenas-Avenda{\~n}o, Masashi Kimura, and Tsvi Piran for helpful comments. DG is supported by a Princeton Gravity Initiative postdoctoral fellowship and by a Princeton Future Faculty in the Physical Sciences fellowship.

\appendix

\section{Frame and sky of a particle orbiting a Kerr BH}  
\label{app:CircularOrbitersFrame}

We will describe the collision in the frame of the circular equatorial NHEK orbiter, particle $\mathrm{I}$, which has Kerr energy $E=m/\sqrt{3}$, azimuthal angular momentum $L=2Mm/\sqrt{3}$, and Carter constant $Q=0$. 
In order to describe the universal kinematics of the collision in that frame, we define the orbiter's frame and emission angles in the subextremal case, before taking the NHEK limit.

A Kerr circular orbiter at radius $r$ obeys $\Theta(\pi/2)=\Theta'(\pi/2)=0$ and $\mathcal{R}(r)=\mathcal{R}'(r)=0$. Solving these equations for its conserved quantities yields $Q=0$, and
\begin{subequations}
\begin{align}
	E&=m\frac{r^{3/2}-2M\sqrt{r}\pm a\sqrt{M}}{\sqrt{r^3-3Mr^2\pm2a\sqrt{M}r^{3/2}}} \,, \\
    L&=\pm m\sqrt{M}\frac{{r^2\mp2a\sqrt{Mr}+a^2}}{\sqrt{r^3-3Mr^2\pm2a\sqrt{M}r^{3/2}}} \,,
\end{align}
\end{subequations}
where the upper/lower sign corresponds to prograde/retrograde orbits, respectively.
These orbiters have four-velocity
\begin{align}
    u=p/m=\frac{\pa{r^{3/2}\pm a\sqrt{M}}\pd_t\pm\sqrt{M}\pd_\phi}{\sqrt{r^3-3Mr^2\pm2a\sqrt{M}r^{3/2}}} \,,
\end{align}
and angular velocity $\Omega=u^\phi/u^t$.
Circular orbits are stable ($\mathcal{R}''(r)<0$) down to the ISCO radius
\begin{subequations}
\begin{align}
\label{eq:ISCO}
	r_{\rm isco}^\pm&=M\br{3+Z_2\mp\sqrt{\pa{3-Z_1}\pa{3+Z_1+2Z_2}}},\\
	Z_1&=1+\sqrt[3]{1-a_\star^2}\pa{\sqrt[3]{1+a_\star}+\sqrt[3]{1-a_\star}},\\
	Z_2&=\sqrt{3a_\star^2+Z_1^2},\quad
	a_\star=a/M.
\end{align}
\end{subequations}

We take the following local orthonormal frame for a circular orbiter
\begin{subequations}
\label{eq:OrthornomalFrame}
\begin{align}
	\mathbf{e}_{[t]}&=u,\\
	\mathbf{e}_{[r]}&=\sqrt{1-\frac{2M}{r}+\frac{a^2}{r^2}}\pd_r,\\
	\mathbf{e}_{[\theta]}&=\frac{1}{r}\pd_\theta,\\
        \mathbf{e}_{[\phi]}&=v u^t\pa{\pd_t+\omega\pd_\phi}+\gamma\sqrt{\frac{\omega r}{2aM}}\pd_\phi,
\end{align}
\end{subequations}
where 
\begin{align}
    \omega=-\frac{g_{t\phi}}{g_{\phi\phi}}=\frac{2aMr}{\pa{r^2+a^2}^2-a^2\Delta} \,,
\end{align} 
is the so-called `frame dragging' angular velocity induced by the BH's rotation, and
\begin{align}
    v=\frac{\pa{r^2+a^2}^2-a^2\Delta}{r^2\sqrt{\Delta}}\pa{\Omega-\omega}, \quad \gamma=\frac{1}{\sqrt{1-v^2}}  \,,
\end{align}
are the velocity and Lorentz factor of the orbiter relative to a locally non-rotating frame.
The frame \eqref{eq:OrthornomalFrame} obeys
\begin{align}
    \label{eq:TetradConditions}
    g_{\mu\nu}\mathbf{e}_{[a]}^\mu\mathbf{e}_{[b]}^\nu=\eta_{[a][b]},\quad
	\eta^{[a][b]}\mathbf{e}_{[a]}^\mu\mathbf{e}_{[b]}^\nu=g^{\mu\nu},
\end{align}
where $\eta^{[a][b]}=\mathrm{diag}\pa{-1,1,1,1}$.
Frame components of four-vectors $V^\mu$ are given by
\begin{align}
	V^{[a]}=\eta^{[a][b]}\mathbf{e}_{[b]}^\mu V_\mu.
\end{align}

Spatial directions at fixed time in the frame of the orbiter, the \emph{orbiter sky}, can be parameterized by angles $\Psi \in \br{0,\pi}$, measured from the orbiter's direction of motion $\mathbf{e}_{[\phi]}$, and $\Upsilon \in \left(-\pi,\pi \right]$, measured from the frame axis parallel to the BH spin axis $\mathbf{e}_{[\theta]}$, in the plane perpendicular to the direction of motion. See Fig.~\ref{fig:OrbiterSkys}.
Emission angles of particles in the orbiter frame are thus given by
\begin{subequations}
\label{eq:SkyAngles}
\begin{align}
    \label{eq:SkyPolarAngle}
    \Psi&=\arccos\frac{p^{[\phi]}}{p^{[t]}} \,, \\
    \label{eq:SkyAzimuthalAngle}
    \Upsilon&= \pm_r\arccos\pa{\frac{1}{\sqrt{1-\cos^2\Psi}} \ \frac{p^{[\theta]}}{p^{[t]}}} \,,
\end{align}  
\end{subequations}
where the orbiter-frame rescaled momenta are
\begin{subequations}
\label{eq:OrbiterFrameMomenta}
\begin{align}
    \frac{p^{[\phi]}}{p^{[t]}}
	&=\frac{1}{1-\Omega\lambda}\br{\frac{\gamma r \lambda}{u^t \sqrt{(r^2+a^2)^2-a^2 \Delta}}-v\pa{1-\omega \lambda}},\\
	\frac{p^{[\theta]}}{p^{[t]}}	&=\pm_\theta\frac{\sqrt{\eta}}{r u^t\pa{1-\Omega \lambda}}.
\end{align}
\end{subequations}
Here, $\lambda=L/E$ and $\eta=Q/E^2$ are the particle's energy-rescaled azimuthal angular momentum and Carter constant. For impinging (opposed to emitted) particles, their direction of arrival to the orbiter is antipoldal to that defined by \eqref{eq:SkyAngles}, i.e. it can be found by taking $\pa{\Psi\mapsto\pi-\Psi,\Upsilon\mapsto\Upsilon+\pi}$ or $\pa{p^{[\phi]}\mapsto-p^{[\phi]},p^{[\theta]}\mapsto-p^{[\theta]}}$.

The sky of any local observer in a single-BH geometry naturally divides into two parts by the behavior of null geodesics intersecting the observer. Photons emitted into the patch of the sky which includes the `direction to the BH center' 
\begin{align}
    \label{eq:BlackHoleCenter}
    \pa{\Psi_\bullet,\Upsilon_\bullet}=\pa{\arccos\pa{-v_s},-\frac{\pi}{2}},
\end{align}
defined by $\lambda=\eta=0$ with $\pm_r=-1$,
will be captured by the BH while those with direction of travel in the complementary patch will escape to infinity. 
Points precisely on the critical curve correspond to a special one-parameter family of $(\lambda,\eta)$ values for which a photon can orbit the BH indefinitely at a fixed radius, known as spherical or (unstably) bound photon orbits \cite{Bardeen1973,Teo2003}.  The spherical photon orbits exist at the \emph{photon shell} region \cite{Johnson2020} given by $\tilde{r}\in\br{\tilde{r}_-,\tilde{r}_+}$ where
 \begin{align}
    \label{eq:PhotonShell}
	\tilde{r}_\pm=2M\br{1+\cos\pa{\frac{2}{3}\arccos\pa{\pm\frac{a}{M}}}}.
\end{align}
The conserved quantities $(\lambda,\eta)$ of the spherical orbits are given by the critical values
\begin{subequations}
\begin{align}
\label{eq:CriticalParameters}
    \tilde{\lambda}(\tilde{r})&=a+\frac{\tilde{r}}{a}\br{\tilde{r}-\frac{2 \Delta(\tilde{r})}{\tilde{r}-M}},\\
    \tilde{\eta}(\tilde{r})&=\frac{\tilde{r}^{3}}{a^2}\br{\frac{4M\Delta(\tilde{r})}{\pa{\tilde{r}-M}^2}-\tilde{r}}.
\end{align}
\end{subequations}
The shape of the critical curve is determined by mapping the locus of 
spherical photon orbits in $(\lambda,\eta)$ space onto the orbiter sky via \eqref{eq:SkyAngles}, \eqref{eq:OrbiterFrameMomenta}, giving the closed curve
\begin{align}
    \label{eq:SkyCriticalCurve}	\mathcal{C}=\cu{\pa{\Psi\pa{\tilde\lambda},\Upsilon\pa{\tilde\lambda,\tilde\eta}}\Big|\tilde{r}_-\le\tilde r\le\tilde{r}_+} \,,
\end{align}
where $\pm_r=\mathrm{sign}\pa{\tilde{r}-r}$ along the critical curve.
Finally, the redshift factor is given by
\begin{align}
    g=\frac{E}{p^{\br{t}}}=\frac{\sqrt{r^3-3Mr^2\pm2a\sqrt{M}r^{3/2}}}{r^{3/2}\pm\sqrt{M}\pa{a-\lambda}}.
\end{align}

\section{Superradiant emission in the sky of a NHEK orbiter}
\label{app:SuperradiantEmission}
Here we elaborate on the relation between the scaling of particles to the superradiant bound, and the angle in which they are emitted or arrive at the NHEK orbiter.
To this end, we consider the triple-scaling limit
\begin{subequations}
\label{eq:DoubleLimit}
\begin{align}
	a&=M\sqrt{1-\kappa^2},
	&&0<\kappa\ll1,\\
	r&=r_+\pa{1+\kappa^p R},
	&&0<p<1,\\
	\lambda&=M\pa{2 + \kappa^q l},
	&&0<q\le 1.
\end{align}
\end{subequations}
 For example, the ISCO is $p=2/3$, $R=2^{1/3}$, as $r_{\mathrm{isco}}^+=r_+\pa{1+2^{1/3}\kappa^{2/3}+\O{\kappa}}$ for $\kappa \ll 1$.
Taking the limit $\kappa\to 0$, we find a relation between the how particles tune to the superradiant bound, and the range of angles with which they can be emitted from the orbiter,
\begin{align}
    &\pa{\cos\Psi,\cos\Upsilon}  \\
    &=\begin{cases}
    \pa{-\dfrac{1}{2},0}, & q<p \,,\\[2ex]
    \pa{\dfrac{l}{3R-2l},\pm_\theta\dfrac{\mathrm{sign}\pa{3R-2l}\sqrt{\eta} R}{M\sqrt{3R^2-4lR+l^2}}}, & q=p \,,\\[2ex]
    \pa{0,\pm_\theta\frac{\sqrt{\eta}}{\sqrt{3}M}},& q>p \,.  \nonumber
    \end{cases}
\end{align}

Thus, particles which tend to the superradiant bound at the same rate as the orbiter radius tends to $r=M$, $\pa{q=p}$, can leave or enter the the orbiter sky from any direction. On the other hand, particles which tend to the superradiant bound faster then the orbiter radius tends to $r=M$, $\pa{q>p}$, can only enter or leave the sky on the curve delineating the forwards and backwards hemispheres. Finally, particles which tend to the superradiant bound at a slower rate than the orbiter radius tends to $r=M$, $\pa{q<p}$---including generic particles---can only enter the sky in the equatorial plane at the two directions with $\Psi=\pi/3$ and leave the sky equatorially at the two directions with $\Psi=2\pi/3$. 

In the case of the collisions considered in this paper, particle I is the NHEK orbiter and particle II is a plunging, `generic' particle, i.e., $\pm_r=-1$ and angular momentum $L\not\approx2M$ in the extremal limit. Therefore, particle II must have $\pa{\Psi_{\mathrm{II}},\Upsilon_{\mathrm{II}}}=\pa{2\pi/3,-\pi/2}$. 
Note that both $\Upsilon=\pm\pi/2$ correspond to generic, non-superradiant particles; however, $\Upsilon=\pi/2$ corresponds to an outgoing particle with $\pm_r=1$, which would need to be emitted by some process deeper in the NHEK. 
This seems like a less probable situation, as most directions in the NHEK orbiter sky correspond to superradiant emission.
The latter fact also implies that 
the image of most of the NHEK orbiter's emission will appear near the NHEKline, for sufficiently inclined observers that have access to it \cite{Gralla2017}.

\section{Bethe-Heitler bremsstrahlung Differential cross section} \label{app: cross section formula}

Here we present the Bethe-Heitler bremsstrahlung differential cross section in the orbiter (particle I) rest frame, when the orbiter is the electron and the plunger is the proton. The reciprocal case is related to the presented one by a boost \cite{Haug2003}.
Defining 
\begin{align}
    |\mathbf{p}|&=\sqrt{\epsilon^2-m_p^2} \,, \\
    X&=\frac{k \pa{\epsilon-|p|\cos\theta}}{m_e} \,,\\
    Y&=\sqrt{|\mathbf{p}|^2+X^2-2\epsilon X} \,, \\
    Z&=\sqrt{Y^2+2k \frac{m_p^2}{m_e}} \,,
\end{align}
the cross section is
\begin{align}
    \label{eq:BHCrossSection}
    \frac{d\sigma}{dk d\Omega}=&\frac{\alpha r_e^2}{2\pi m_e}\Bigg\{ \frac{Y\pa{|\mathbf{p}|-\epsilon\cos\theta}}{|\mathbf{p}|^2} +\frac{Y\br{\pa{k-m_e}|\mathbf{p}|-k\epsilon\cos\theta}}{kZ^2}\nonumber\\
    &+\frac{m_eY}{k|\mathbf{p}|^3}\br{\epsilon^2+\frac{k m_p^2\epsilon}{X m_e}-2\pa{2\epsilon^2+m_p^2}\cos^2\theta}\nonumber\\
    &-\frac{2m_p^2}{|\mathbf{p}|X}\ln\br{\frac{\epsilon+Y-X}{m_p}}+\frac{A m_p^2}
    {k|\mathbf{p}|m_e Z^3} \ln\br{\frac{Z+Y}{Z-Y}} \nonumber\\
    &+\frac{B}{k^3|\mathbf{p}|^4}\ln\br{\frac{ |\mathbf{p}|\pa{|\mathbf{p}|+Y}}{m_p X}-\frac{\epsilon}{m_p}} \Bigg\} \,,
\end{align}
where
\begin{align}
    A=&-3k^2 m_p^2+k m_e\br{4m_p^2-\pa{X-2\epsilon}\pa{X-\epsilon}}+m_e^2 Y^2\,, \\
    B=& - k^3m_p^2\br{m_p^2+\pa{X-2\epsilon}\epsilon}\nonumber\\
    &-k^2m_em_p^2\br{|\mathbf{p}|^2-X\pa{X+\epsilon-6\epsilon\cos^2\theta}}\nonumber\\
    &+2k^2m_e\br{\epsilon^3\pa{\epsilon-X}+|\mathbf{p}|^2\pa{X^2-X\epsilon+\epsilon^2}}\nonumber\\
    &+4 k m_e^2\epsilon^2X\pa{X-\epsilon}-2m_e^3X^2\epsilon\pa{X-\epsilon} \,,
\end{align}
$\alpha$ is the fine structure constant, and $r_e$ is the classical electron radius.
Note that this cross section has an emission direction cutoff $\theta_{\mathrm{max}}(k)$ \eqref{eq:CosThetaMaxPEB}, which does not appear in the no-recoil approximation but can be enforced in evaluations of the emission properties of Sec.~\ref{sec:escape general properties} by taking $d\sigma/(dk d\Omega) \propto \Theta\br{\cos\theta-\cos\theta_\mathrm{max}(k)}$, as described in Sec.~\ref{sec:brem}. 
Lastly, in the high-energy limit $\epsilon\gg m_p$, the cross section \eqref{eq:BHCrossSection} is, to leading order,
\begin{align}
    \frac{d\sigma}{dk d\Omega}=&\frac{\alpha r_e^2}{\pi m_e}\ln\pa{\frac{\epsilon}{m_p}} C \nonumber\\
    C=&\pa{1-\cos\theta}^2\frac{k}{m_e} -\pa{1+\cos^2\theta}\pa{1-\cos\theta}\nonumber\\
    &+\pa{1+\cos^2\theta}\frac{m_e}{k}.
\end{align}

\bibliographystyle{utphys2}
\bibliography{brems.bib}

\providecommand{\href}[2]{#2}\begingroup\raggedright\begin{thebibliography}{10}

\bibitem{Piran1975}
T.~{Piran}, J.~{Shaham}, and J.~{Katz}, ``{High Efficiency of the Penrose
  Mechanism for Particle Collisions},''
  \href{http://dx.doi.org/10.1086/181755}{{\em \apjl} {\bfseries 196} (Mar.,
  1975) L107}.

\bibitem{Banados2009}
M.~{Ba{\~n}ados}, J.~{Silk}, and S.~M. {West}, ``{Kerr Black Holes as Particle
  Accelerators to Arbitrarily High Energy},''
  \href{http://dx.doi.org/10.1103/PhysRevLett.103.111102}{{\em \prl} {\bfseries
  103} no.~11, (Sept., 2009) 111102},
  \href{http://arxiv.org/abs/0909.0169}{{\ttfamily arXiv:0909.0169 [hep-ph]}}.

\bibitem{Harada2011}
T.~{Harada} and M.~{Kimura}, ``{Collision of an innermost stable circular orbit
  particle around a Kerr black hole},''
  \href{http://dx.doi.org/10.1103/PhysRevD.83.024002}{{\em \prd} {\bfseries 83}
  no.~2, (Jan., 2011) 024002}, \href{http://arxiv.org/abs/1010.0962}{{\ttfamily
  arXiv:1010.0962 [gr-qc]}}.

\bibitem{Reynolds2019}
C.~S. {Reynolds}, ``{Observing black holes spin},''
  \href{http://dx.doi.org/10.1038/s41550-018-0665-z}{{\em Nature Astronomy}
  {\bfseries 3} (Jan., 2019) 41--47},
  \href{http://arxiv.org/abs/1903.11704}{{\ttfamily arXiv:1903.11704
  [astro-ph.HE]}}.

\bibitem{Draghis2023}
P.~A. Draghis, J.~M. Miller, M.~C. Brumback, A.~C. Fabian, J.~A. Tomsick, and
  A.~Zoghbi, ``{An Extreme Black Hole in the Recurrent X-ray Transient XTE
  J2012+381},'' \href{http://arxiv.org/abs/2307.06988}{{\ttfamily
  arXiv:2307.06988 [astro-ph.HE]}}.

\bibitem{Piran1975a}
T.~{Piran} and J.~{Shaham}, ``{Can soft {\ensuremath{\gamma}}-ray bursts be
  emitted by accreting black holes?},''
  \href{http://dx.doi.org/10.1038/256112a0}{{\em \nat} {\bfseries 256}
  no.~5513, (July, 1975) 112--113}.

\bibitem{Piran1977}
T.~{Piran} and J.~{Shaham}, ``{Production of gamma-ray bursts near rapid
  rotating accreting black holes.},''
  \href{http://dx.doi.org/10.1086/155251}{{\em \apj} {\bfseries 214} (May,
  1977) 268--299}.

\bibitem{Piran1977a}
T.~{Piran} and J.~{Shaham}, ``{Upper bounds on collisional Penrose processes
  near rotating black-hole horizons},''
  \href{http://dx.doi.org/10.1103/PhysRevD.16.1615}{{\em \prd} {\bfseries 16}
  no.~6, (Sept., 1977) 1615--1635}.

\bibitem{Banados2011}
M.~{Ba{\~n}ados}, B.~{Hassanain}, J.~{Silk}, and S.~M. {West}, ``{Emergent flux
  from particle collisions near a Kerr black hole},''
  \href{http://dx.doi.org/10.1103/PhysRevD.83.023004}{{\em \prd} {\bfseries 83}
  no.~2, (Jan., 2011) 023004}, \href{http://arxiv.org/abs/1010.2724}{{\ttfamily
  arXiv:1010.2724 [astro-ph.CO]}}.

\bibitem{Bardeen1999}
J.~{Bardeen} and G.~T. {Horowitz}, ``{Extreme Kerr throat geometry: A vacuum
  analog of AdS$_{2}\times$S$^{2}$},''
  \href{http://dx.doi.org/10.1103/PhysRevD.60.104030}{{\em \prd} {\bfseries 60}
  no.~10, (Nov., 1999) 104030},
  \href{http://arxiv.org/abs/hep-th/9905099}{{\ttfamily arXiv:hep-th/9905099
  [hep-th]}}.

\bibitem{Guica2009}
M.~{Guica}, T.~{Hartman}, W.~{Song}, and A.~{Strominger}, ``{The Kerr/CFT
  correspondence},'' \href{http://dx.doi.org/10.1103/PhysRevD.80.124008}{{\em
  \prd} {\bfseries 80} no.~12, (Dec., 2009) 124008},
  \href{http://arxiv.org/abs/0809.4266}{{\ttfamily arXiv:0809.4266 [hep-th]}}.

\bibitem{Compere2012}
G.~{Comp{\`e}re}, ``{The Kerr/CFT correspondence and its extensions: a
  comprehensive review},''
  \href{http://dx.doi.org/10.48550/arXiv.1203.3561}{{\em arXiv e-prints} (Mar.,
  2012) arXiv:1203.3561}, \href{http://arxiv.org/abs/1203.3561}{{\ttfamily
  arXiv:1203.3561 [hep-th]}}.

\bibitem{Hadar2014}
S.~{Hadar}, A.~P. {Porfyriadis}, and A.~{Strominger}, ``{Gravity waves from
  extreme-mass-ratio plunges into Kerr black holes},''
  \href{http://dx.doi.org/10.1103/PhysRevD.90.064045}{{\em \prd} {\bfseries 90}
  no.~6, (Sept., 2014) 064045},
  \href{http://arxiv.org/abs/1403.2797}{{\ttfamily arXiv:1403.2797 [hep-th]}}.

\bibitem{Gralla2017}
S.~E. {Gralla}, A.~{Lupsasca}, and A.~{Strominger}, ``{Observational signature
  of high spin at the Event Horizon Telescope},''
  \href{http://dx.doi.org/10.1093/mnras/sty039}{{\em \mnras} {\bfseries 475}
  no.~3, (Apr, 2018) 3829--3853},
  \href{http://arxiv.org/abs/1710.11112}{{\ttfamily arXiv:1710.11112
  [astro-ph.HE]}}.

\bibitem{Compere2017}
G.~{Comp{\`e}re} and R.~{Oliveri}, ``{Self-similar accretion in thin discs
  around near-extremal black holes},''
  \href{http://dx.doi.org/10.1093/mnras/stx748}{{\em \mnras} {\bfseries 468}
  no.~4, (July, 2017) 4351--4361},
  \href{http://arxiv.org/abs/1703.00022}{{\ttfamily arXiv:1703.00022
  [astro-ph.HE]}}.

\bibitem{Gates2018}
D.~{Gates}, D.~{Kapec}, A.~{Lupsasca}, Y.~{Shi}, and A.~{Strominger},
  ``{Polarization Whorls from M87 at the Event Horizon Telescope},''
  \href{http://dx.doi.org/10.1098/rspa.2019.0618}{{\em Proceedings of the Royal
  Society A} {\bfseries 476} no.~2237, (Sep, 2020) 20190618},
  \href{http://arxiv.org/abs/1809.09092}{{\ttfamily arXiv:1809.09092
  [hep-th]}}.

\bibitem{Gates2021}
D.~E.~A. {Gates}, S.~{Hadar}, and A.~{Lupsasca}, ``{Photon emission from
  circular equatorial Kerr orbiters},''
  \href{http://dx.doi.org/10.1103/PhysRevD.103.044050}{{\em \prd} {\bfseries
  103} no.~4, (Feb., 2021) 044050},
  \href{http://arxiv.org/abs/2010.07330}{{\ttfamily arXiv:2010.07330 [gr-qc]}}.

\bibitem{Bejger2012}
M.~{Bejger}, T.~{Piran}, M.~{Abramowicz}, and F.~{H{\r{a}}kanson},
  ``{Collisional Penrose Process near the Horizon of Extreme Kerr Black
  Holes},'' \href{http://dx.doi.org/10.1103/PhysRevLett.109.121101}{{\em \prl}
  {\bfseries 109} no.~12, (Sept., 2012) 121101},
  \href{http://arxiv.org/abs/1205.4350}{{\ttfamily arXiv:1205.4350
  [astro-ph.HE]}}.

\bibitem{McWilliams2013}
S.~T. {McWilliams}, ``{Black Holes are Neither Particle Accelerators Nor Dark
  Matter Probes},''
  \href{http://dx.doi.org/10.1103/PhysRevLett.110.011102}{{\em \prl} {\bfseries
  110} no.~1, (Jan., 2013) 011102},
  \href{http://arxiv.org/abs/1212.1235}{{\ttfamily arXiv:1212.1235 [gr-qc]}}.

\bibitem{Harada2014}
T.~{Harada} and M.~{Kimura}, ``{Black holes as particle accelerators: a brief
  review},'' \href{http://dx.doi.org/10.1088/0264-9381/31/24/243001}{{\em
  Classical and Quantum Gravity} {\bfseries 31} no.~24, (Dec., 2014) 243001},
  \href{http://arxiv.org/abs/1409.7502}{{\ttfamily arXiv:1409.7502 [gr-qc]}}.

\bibitem{Leiderschneider2016}
E.~{Leiderschneider} and T.~{Piran}, ``{Maximal efficiency of the collisional
  Penrose process},'' \href{http://dx.doi.org/10.1103/PhysRevD.93.043015}{{\em
  \prd} {\bfseries 93} no.~4, (Feb., 2016) 043015},
  \href{http://arxiv.org/abs/1510.06764}{{\ttfamily arXiv:1510.06764 [gr-qc]}}.

\bibitem{Thorne1974}
K.~S. {Thorne}, ``{Disk-Accretion onto a Black Hole. II. Evolution of the
  Hole},'' \href{http://dx.doi.org/10.1086/152991}{{\em \apj} {\bfseries 191}
  (July, 1974) 507--520}.

\bibitem{Berti2009}
E.~{Berti}, V.~{Cardoso}, L.~{Gualtieri}, F.~{Pretorius}, and U.~{Sperhake},
  ``{Comment on ``Kerr Black Holes as Particle Accelerators to Arbitrarily High
  Energy''},'' \href{http://dx.doi.org/10.1103/PhysRevLett.103.239001}{{\em
  \prl} {\bfseries 103} no.~23, (Dec., 2009) 239001},
  \href{http://arxiv.org/abs/0911.2243}{{\ttfamily arXiv:0911.2243 [gr-qc]}}.

\bibitem{Gralla2016}
S.~E. {Gralla}, A.~{Lupsasca}, and A.~{Strominger}, ``{Near-horizon Kerr
  magnetosphere},'' \href{http://dx.doi.org/10.1103/PhysRevD.93.104041}{{\em
  \prd} {\bfseries 93} no.~10, (May, 2016) 104041},
  \href{http://arxiv.org/abs/1602.01833}{{\ttfamily arXiv:1602.01833
  [hep-th]}}.

\bibitem{Bredberg2010}
I.~{Bredberg}, T.~{Hartman}, W.~{Song}, and A.~{Strominger}, ``{Black hole
  superradiance from Kerr/CFT},''
  \href{http://dx.doi.org/10.1007/JHEP04(2010)019}{{\em Journal of High Energy
  Physics} {\bfseries 2010} (Apr., 2010) 19},
  \href{http://arxiv.org/abs/0907.3477}{{\ttfamily arXiv:0907.3477 [hep-th]}}.

\bibitem{Bethe1934}
H.~{Bethe} and W.~{Heitler}, ``{On the Stopping of Fast Particles and on the
  Creation of Positive Electrons},''
  \href{http://dx.doi.org/10.1098/rspa.1934.0140}{{\em Proceedings of the Royal
  Society of London Series A} {\bfseries 146} no.~856, (Aug., 1934) 83--112}.

\bibitem{Sauter1934}
F.~{Sauter}, ``{{\"U}ber die Bremsstrahlung schneller Elektronen},''
  \href{http://dx.doi.org/10.1002/andp.19344120405}{{\em Annalen der Physik}
  {\bfseries 412} no.~4, (Jan., 1934) 404--412}.

\bibitem{Koch1959}
H.~W. {Koch} and J.~W. {Motz}, ``{Bremsstrahlung Cross-Section Formulas and
  Related Data},'' \href{http://dx.doi.org/10.1103/RevModPhys.31.920}{{\em
  Reviews of Modern Physics} {\bfseries 31} no.~4, (Oct., 1959) 920--955}.

\bibitem{Haug2003}
E.~{Haug}, ``{Proton-electron bremsstrahlung},''
  \href{http://dx.doi.org/10.1051/0004-6361:20030782}{{\em \aap} {\bfseries
  406} (July, 2003) 31--35}.

\bibitem{Bardeen1973}
J.~M. {Bardeen}, ``{Timelike and null geodesics in the Kerr metric.},'' in {\em
  Black Holes (Les Astres Occlus)}, C.~{Dewitt} and B.~S. {Dewitt}, eds.,
  pp.~215--239.
\newblock Gordon and Breach Science Publishers, Jan, 1973.

\bibitem{Teo2003}
E.~{Teo}, ``{Spherical Photon Orbits Around a Kerr Black Hole},''
  \href{http://dx.doi.org/10.1023/A:1026286607562}{{\em General Relativity and
  Gravitation} {\bfseries 35} no.~11, (Nov, 2003) 1909--1926}.

\bibitem{Johnson2020}
M.~D. {Johnson}, A.~{Lupsasca}, A.~{Strominger}, G.~N. {Wong}, S.~{Hadar},
  D.~{Kapec}, R.~{Narayan}, A.~{Chael}, C.~F. {Gammie}, P.~{Galison}, D.~C.~M.
  {Palumbo}, S.~S. {Doeleman}, L.~{Blackburn}, M.~{Wielgus}, D.~W. {Pesce},
  J.~R. {Farah}, and J.~M. {Moran}, ``{Universal interferometric signatures of
  a black hole's photon ring},''
  \href{http://dx.doi.org/10.1126/sciadv.aaz1310}{{\em Science Advances}
  {\bfseries 6} no.~12, (Mar., 2020) eaaz1310},
  \href{http://arxiv.org/abs/1907.04329}{{\ttfamily arXiv:1907.04329
  [astro-ph.IM]}}.

\end{thebibliography}\endgroup

\end{document}